\documentclass[9pt,conference]{IEEEtran}
\usepackage[preprint]{dcase2026}

\usepackage{bm} 
\usepackage{comment}

\usepackage{dcase2026,amsmath,graphicx,url,times,booktabs, tabularx}
\usepackage{csquotes}

\title{Efficient Text-to-Audio Generation via Pruning}

\name{Arshdeep Singh$^{1}$,
      Yi Yuan$^{2}$,
      Yun Chen$^{1}$,
      Wenwu Wang$^{2}$,
      Mark D. Plumbley$^{1}$}
\address{$^{1}$King's College London (KCL),\;
$^{2}$University of Surrey, UK.
}

\begin{document}

\maketitle

\begin{abstract}
Diffusion-based text-to-audio generative models such as AudioLDM achieve high perceptual quality and strong semantic consistency; however, their practical deployment is hindered by the substantial computational cost of the U-Net denoising backbone. In this work, we apply model pruning to improve the computational efficiency of AudioLDM, a U-Net based text-conditioned audio latent diffusion model. We analyse parameter redundancy across U-Net convolutional blocks and evaluate a filter-pruning strategy. Pruning is guided by norm-based criteria and followed by lightweight finetuning to recover performance losses. Experimental results demonstrate that up to 83\% of the parameters and 39\% of the multiply–accumulate operations of U-Net have been reduced while maintaining, and in some cases improving, generation quality compared to the baseline unpruned network. We find that pruning affects AudioLDM’s ability to generate certain sound events including safety-critical sounds such as gunshots, sirens, and explosions, as well as mechanical sounds such as drills and sewing machines, and other sounds such as sprays and tick-tocks, which are mostly recovered  by lightweight finetuning of the pruned model.
\end{abstract}

\begin{IEEEkeywords}
Diffusion model, TTA, efficiency, AudioLDM, pruning.
\end{IEEEkeywords}

\section{Introduction}
\label{sec:intro}

The text-to-audio generation \cite{huang2023make} task aims to synthesize audio signals directly from textual descriptions. Given a natural language input such as ``a dog barking in a park'' or ``soft piano music with rain in the background,'' a text-to-audio generative model produces an audio waveform that semantically matches the description. 




Recent progress in generative modelling has led to the widespread use of diffusion-based approaches, such as AudioLDM \cite{liu2023audioldm} for text-to-audio generation. Diffusion models generate data by gradually transforming noise into structured signals through a sequence of iterative denoising steps. During this process, the model learns to remove noise from intermediate representations while being conditioned on text embeddings. 

Despite their success \cite{yuan2026dreamaudio}, diffusion-based text-to-audio models suffer from significant efficiency challenges \cite{ma2025efficient}. 
Each generated sample in a diffusion model requires many sequential denoising steps and uses the denoising network repeatedly; therefore, energy usage scales with the number of sampling steps.  On the other hand, increasing the number of steps generally improves generation quality, as the model has more opportunities to refine the output. Therefore, there is a trade-off between generation quality and efficiency.

Efficiency in audio synthesis models is achieved through foundational strategies adapted from visual domains, such as latent space modelling used in models like Make-an-Audio \cite{huang2023make} and DiffSound \cite{yang2023diffsound} to remove signal redundancy and reduce the high computational overhead of processing raw waveforms. Accelerated sampling strategies, including the application of specialized stochastic or ordinary differential equation solvers \cite{chung2022come}, further improve inference speeds by  decreasing the required number of iterative denoising steps \cite{kongdiffwave}. Efficiency is further enhanced through knowledge distillation \cite{hinton2015distilling}, where a \enquote{student} model learns to approximate the denoising trajectory of a \enquote{teacher} model in a single or significantly
fewer steps \cite{salimansprogressive}. Also, lightweight training using LoRA techniques \cite{niu2024soundlocd} has been utilized to update fewer parameters during training.

\textcolor{black}{Existing   methods for audio-based diffusion models  do not explicitly consider computational efficiency of the diffusion model itself. Typically, diffusion models are large-scale models with millions of parameters that require significant memory and have high multiply-accumulate operations (MACs) to compute output. This makes such models resource-hungry, and  also  the denoising process is repeated multiple times during generation, which makes large-scale diffusion models to consume significant resources.}



To address the complexity of large-scale models, convolutional filter pruning \cite{he2023structured}  approaches have been utilized. These techniques  eliminate redundant or unimportant parameters based on their importance \cite{singh2025efficient}. The filter importance quantifies how well the underlying filter contributes to model performance. Removing less important parameters reduces memory usage and computational complexity, while lightweight finetuning helps preserve model performance.

Majority of the existing pruning methods applied to diffusion model such as  LD-Pruner \cite{castells2024ld}, LAPTOP-Diff \cite{zhang2024laptop}, LayerMerge \cite{kim2024layermerge}, and Diff-Pruning \cite{fang2023structural} 
have been applied in vision domain. These methods use datasets to calculate importance scores for model compression. For instance, LD-Pruner \cite{castells2024ld} employs a task-agnostic importance metric that operates in the latent space, using input conditions or prompts to identify which operators or filters to remove. 
However, the process of evaluating a model's performance to determine pruning importance is a cumbersome and resource-intensive task. This is particularly challenging for generative models where standard metrics such as Fr\'echet Inception Distance (FID) require generating thousands of images, which can require more than an hour for just a single evaluation. Furthermore,  constructing look-up tables to quantify importance  using a full training dataset may consume up to 126 GPU hours \cite{kim2024layermerge}.  Additionally, the pruned model might suffer  from overfitting, where the pruned model might only perform well on the specific prompts or conditions present in the calibration set,  requiring  diverse datasets or multiple prompts to maintain robustness \cite{castells2024ld}.


In this work, we leverage a filter pruning method to improve the efficiency of diffusion-based text-to-audio generation model, AudioLDM. Our goal is to estimate filter importance using only the diffusion model parameters, without relying on any external dataset, thereby reducing the computational cost associated with importance estimation, followed by finetuning the pruned model to recover its performance loss.
We reduce 83\% of the total parameter count and 39\% MACs of U-Net while maintaining similar audio quality compared to that of unpruned AudioLDM model. We also explore the effect of pruning on generation quality, and find that pruning degrades the generation quality of sound events including safety-critical events such as gunshots, sirens, explosions and mechanical sounds such as drill, sewing machine or sounds like spray, tick-tock, which are then mostly recovered  by lightweight finetuning of the pruned model. Our implementation is available on \url{https://github.com/Arshdeep-Singh-Boparai/PruningAudioLDM.git}.

\section{Background and related work}

\subsection{AudioLDM architecture overview}

\begin{figure}
    \centering
    \includegraphics[scale=0.457]{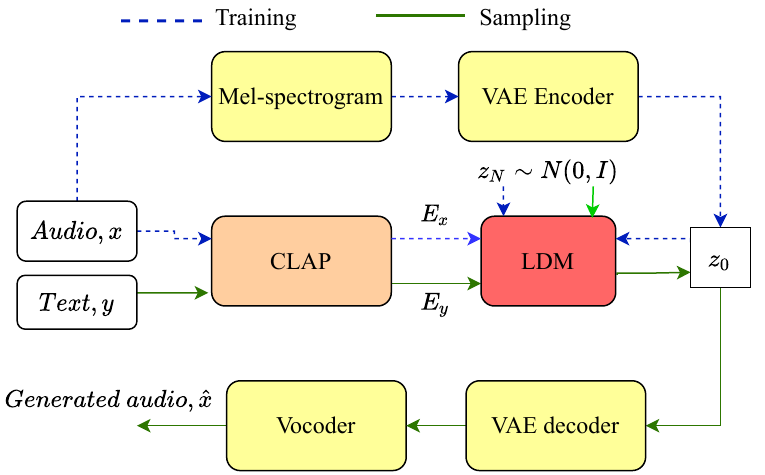}
    \vspace{-0.24cm}
    \caption{An overview of the AudioLDM pipeline.}
    \label{fig:audioldm}
\end{figure}

AudioLDM \cite{liu2023audioldm} is a latent diffusion-based text-to-audio (TTA) generation model that leverages contrastive language-audio pretraining (CLAP) embeddings to align text and audio in a shared representation space. As shown in Fig.~\ref{fig:audioldm}, it consists of a CLAP encoder, a variational autoencoder (VAE), a latent diffusion model (LDM), and a HiFi-GAN vocoder. The CLAP model extracts aligned text and audio embeddings, denoted by $E_y$ and $E_x$, respectively. The VAE compresses mel-spectrograms into a compact latent representation $z_0$, while the LDM learns the conditional distribution $q(z_0|E_y)$ in the latent space. During inference, the generated latent representation is decoded by the VAE and converted into the final audio waveform using the vocoder.

\subsection{U-Net based LDM architecture in AudioLDM}

The LDM has a U-Net architecture with 
 four encoder blocks, a middle block, and
four decoder blocks. With a basic channel number of $c_u$, the channel dimensions of encoder blocks are $[b_1c_u,b_2c_u,b_3c_u,b_4c_u]$ with $(b_1,b_2,b_3,b_4) \in \mathbb{Z_{+}}$, are the \enquote{channel scaling parameters} for different blocks. The channel dimensions of the decoder blocks are the reverse of those of the encoder blocks, and the channel of the middle block has 5$c_u$
dimensions.

\subsection{Filter pruning}

Filter pruning methods \cite{he2023structured}  reduce model size, memory requirements, and computational complexity by removing less important filters that contribute least to model performance. Existing filter pruning approaches can be broadly categorized into \emph{active} and \emph{passive} methods \cite{singh2025efficient}. Active methods estimate filter importance using training data and feature maps, employing metrics such as entropy \cite{li2019using}, rank \cite{lin2020hrank}, or energy \cite{lin2022ezcrop}, but incur significant computational and memory overhead. In contrast, passive methods \cite{singh2025efficient} are data-free and rely solely on filter weights, typically using criteria such as the $\ell_1$-norm \cite{li2017pruning} or filter similarity to identify redundant filters \cite{king2023compressing}. After removing  filters from  the original unpruned network, the pruned network is finetuned to recover most of the performance loss due to pruning.


\section{Efficient AudioLDM with U-Net pruning}

In this work, we apply a  passive filter pruning method on an unpruned U-Net model within AudioLDM. First, we analyse the number of parameters and MACs across various blocks of the U-Net. Next, we apply $l_1$-norm based filter pruning  across  those blocks of the U-Net which contribute significantly to the total parameter count and computational cost. 


\textbf{Block-wise resource analysis:} 
We analyse the number of parameters and computational cost by varying channel scaling parameters   across different blocks in the unpruned U-Net, which has $(b_1,b_2,b_3,b_4)$ = $(1,2,3,5)$, $c_u$ = 192 with 416M parameter count and  a computational complexity of 102.94G MACs.
In Table~\ref{tab:unet_scaling}, we observe that  the deeper blocks of the U-Net contribute disproportionately to the model size. In particular, reducing the channel scaling parameter $b_4$ from 5 to 1 decreases the number of parameters from 416M to 145M, corresponding to a reduction of approximately 65\%, while the computational complexity decreases from 102.94G to 83.79G. Similarly, reducing $b_3$ from 3 to 1 lowers the parameter count from 416M to 331M and the computational complexity from 102.94G to 80.97G. In contrast, modifying $b_2$ yields a comparatively smaller reduction in parameters, decreasing the model size by only 30M parameters. 

The analysis reveals that the majority of U-Net parameters are concentrated in the deeper network blocks, $b_3$ and $b_4$. Therefore, pruning filters from these blocks offers the highest potential for parameter reduction while maintaining the overall network structure. Based on this observation, we apply a pruning framework to the convolutional layers within the $b_3$ and $b_4$ blocks.




\textbf{$l_1$-norm based filter pruning:} After identifying the U-Net blocks targeted for pruning, the importance of convolutional filters is evaluated on a layer-wise basis to determine whether a filter should be retained or removed. For each convolutional layer, the importance of a convolutional filter $\mathbf{F}$ is quantified using its $l_1$-norm, defined as

\begin{equation}
||\mathbf{F}||_1 = \sum_{i=1}^{\text{length}(\mathbf{F})} |\mathbf{F}_i|.
\end{equation}

\begin{figure}[h]
    \centering
    \includegraphics[scale=0.45]{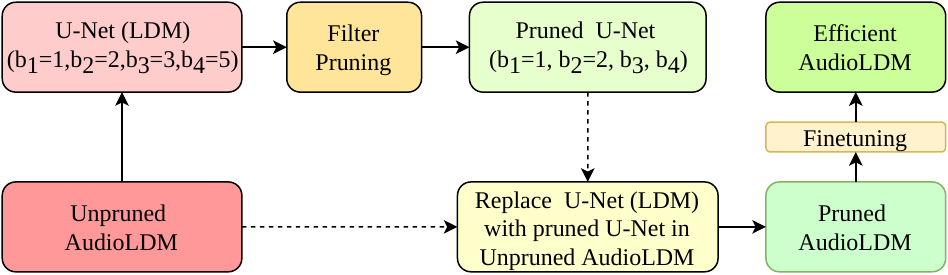}
    \vspace{-0.24cm}
    \caption{An overall framework to obtain efficient AudioLDM.}
    \label{fig:overall framework}
\end{figure}

The filters are subsequently ranked in descending order according to their $l_1$-norm values. For a given convolutional layer, filters with lower $l_1$-norm values are pruned, as they typically generate weaker activations compared to filters with larger norms. Consequently, these filters contribute less to the representational capacity of the model, and their removal may have a minimal impact on model performance  \cite{li2017pruning}. Following the pruning stage, the resulting pruned U-Net is finetuned to recover any performance degradation incurred during the pruning process. An overall framework to obtain efficient AudioLDM is shown in Fig. \ref{fig:overall framework}

\begin{table}[t]
\centering
\caption{Parameters and MACs obtained by scaling individual U-Net blocks while keeping the remaining blocks fixed at the baseline configuration $(b_1,b_2,b_3,b_4)=(1,2,3,5)$ and $c_u=192$.}
\label{tab:unet_scaling}

\resizebox{0.7\columnwidth}{!}{%
\begin{tabular}{c c c c}
\hline
Scaled Block & Value & Params (M) & MACs (G) \\
\hline
$b_4$ & 1 & 145 & 83.79 \\
      & 2 & 182 & 86.40 \\
      & 4 & 317 & 95.98 \\
      & 5 & 416 & 102.94 \\
\hline
$b_3$ & 1 & 331 & 80.97 \\
      & 2 & 365 & 89.63 \\
      & 3 & 416 & 102.94 \\
\hline
$b_2$ & 1 & 385 & 71.50 \\
      & 2 & 416 & 102.94 \\
\hline
\end{tabular}%
}
\end{table}

\section{Experimental setup}

\textbf{Dataset used:} We use  AudioCaps dataset \cite{kim2019audiocaps} for our experiments. AudioCaps is a large-scale audio-text paired dataset consisting of audio clips and human-written captions collected through crowdsourcing from the AudioSet dataset. It contains approximately 49,000 audio–text pairs for training and 964 audio–text pairs for testing. In addition, AudioCaps provides event-level ground-truth annotations for each audio clip, indicating the audio events present in each recording.

\noindent \textbf{Unpruned AudioLDM baseline:} 
We use a pre-trained  AudioLDM-M-Full \cite{liu2023audioldm}, as a baseline unpruned model. The LDM in AudioLDM-M-Full has a U-Net architecture with \(c_u = 192\) and \((b_1,b_2,b_3,b_4)=(1,2,3,5)\). The LDM has a total of 416M parameters and requires 103G MACs.

AudioLDM-M-Full is trained on the AudioSet \cite{gemmeke2017audio}, AudioCaps \cite{kim2019audiocaps}, FreeSound\footnote{https://freesound.org/}, and BBC SFX datasets\footnote{https://sound-effects.bbcrewind.co.uk/search}, using 3.1M 10-second audio samples for 1.5M training steps. The model is further finetuned on the AudioCaps dataset for an additional 0.25M steps. More details about the training procedure of AudioLDM-M-Full can be found in \cite{liu2023audioldm}.


\noindent \textbf{Obtaining pruned AudioLDM-M-Full and finetuning:} 
We prune the U-Net based LDM by removing filters with smaller \(l_1\)-norms from convolutional layers in the deeper blocks of the unpruned U-Net.

The number of filters removed from the unpruned U-Net are determined by  channel scaling parameters (\(b_3\) and \(b_4\)) in the deeper blocks. Specifically, we select \(b_3 \in \{1, 2\}\) and \(b_4 \in \{1, 2, 4\}\) to obtain different pruned versions of AudioLDM-M-Full.

The pruned AudioLDM-M-Full models are finetuned on the AudioCaps training dataset for 1M steps. 
During finetuning, we update only U-Net model parameters and other components such as VAE, CLAP in AudioLDM pipeline are kept frozen. We follow the same finetuning configuration as used for training  AudioLDM-M-Full model \cite{liu2023audioldm}, except for the training dataset, which is performed on AudioCaps training dataset. The finetuned pruned models are then evaluated on the AudioCaps test dataset.


\noindent \textbf{Models used for comparison:} 
We use DiffSound \cite{yang2023diffsound}, AudioGen \cite{kreuk2022audiogen}, AudioLDM-S/L-Full \cite{liu2023audioldm} and AudioLDM2 \cite{liu2024audioldm} to compare with pruned AudioLDM-M-Full. DiffSound \cite{yang2023diffsound} is a text‑to‑audio generation framework that employs a discrete diffusion model to synthesize mel‑spectrogram tokens conditioned on textual input. AudioGen \cite{kreuk2022audiogen} is an autoregressive Transformer‑based model designed for text‑conditioned audio generation using discrete waveform tokens. Also, we use other variant of AudioLDM models including AudioLDM-S-Full, AudioLDM-L-Full, and AudioLDM2-Full-Large \cite{liu2024audioldm}. For a fair comparison, we also apply finetuning to the unpruned baseline model and evaluate its performance relative to that of the pruned models.


\noindent \textbf{Evaluation metrics:}  We use Fr\'echet Audio Distance (FAD), Kullback-Leibler(KL) divergence performance metrics to evaluate generation quality. The number of inference steps used in generating an audio sample is 200. We compare number of parameters, MACs for analysing computational efficiency across pruned models. Additionally, we report real-time factor (RTF) for generating a text-conditioned 10-seconds audio on NVIDIA GeForce RTX 3090 with 200 inference steps, and memory storage required for AudioLDM checkpoints. For fair comparison with other existing models \cite{yang2023diffsound,kreuk2022audiogen, liu2023audioldm, liu2024audioldm}, we compare only the number of parameters, as the MACs of these models are not publicly available.

\section{Results and Analysis}

\begin{figure}
    \centering
    \includegraphics[scale=0.273]{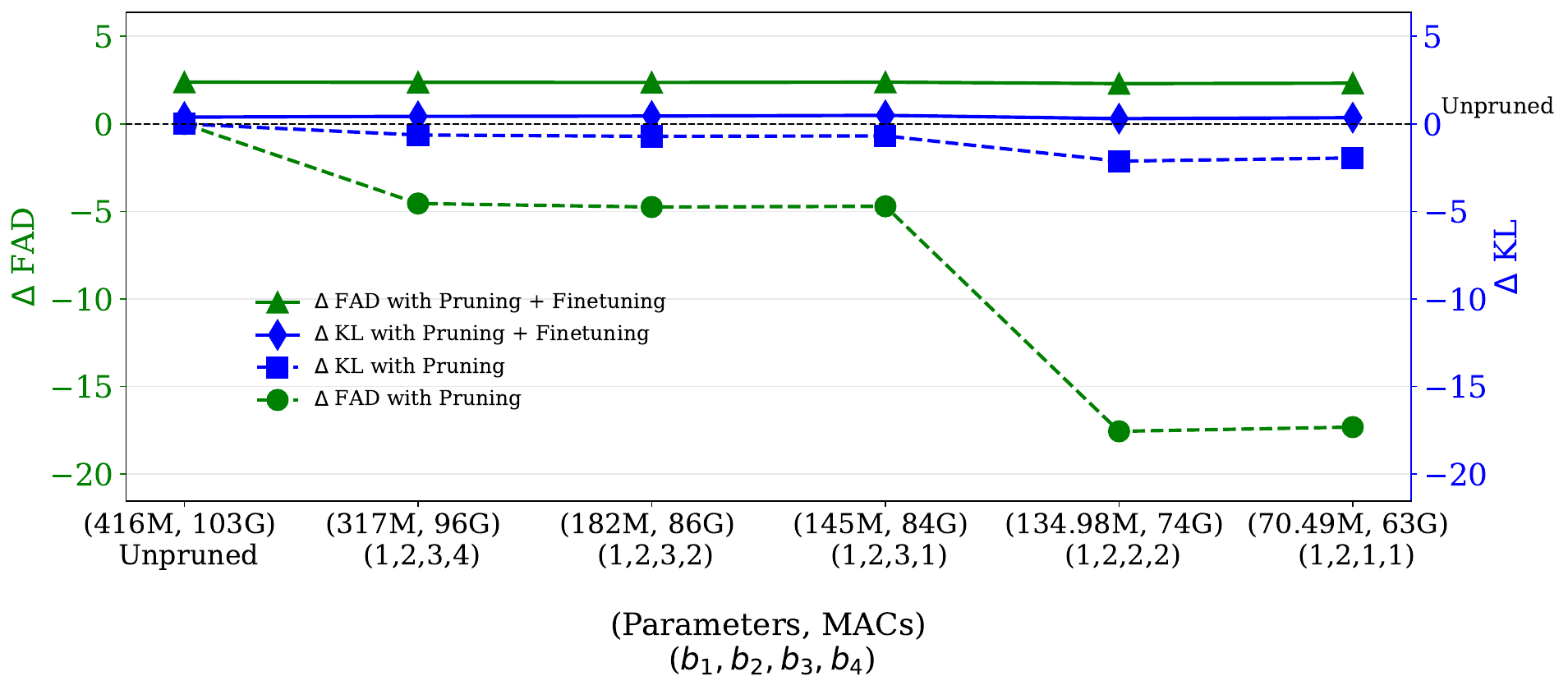}
    \vspace{-0.65cm}
    \caption{Absolute change in FAD and KL of pruned models relative to the unpruned baseline model after applying pruning with and without any finetuning at different $(b_1,b_2,b_3,b_4)$ configurations within U-Net. Parameters count  and MACs are also shown.}
    \label{fig: performance comparison baseline, pruned, pruned/finetuned}
\end{figure}

We compare the performance of the pruned models, both with and without finetuning, with that of the unpruned baseline in Fig.~\ref{fig: performance comparison baseline, pruned, pruned/finetuned}. The unpruned AudioLDM-M-Full model achieves an FAD score of 3.95 and a KL divergence of 2.16 on the AudioCaps evaluation dataset. Pruning leads to a degradation in performance relative to the unpruned model, with the extent of degradation generally increasing as the model size is reduced. 

On the other hand, finetuning recovers most of the performance lost due to pruning and, in some cases, yields results better than that of the unpruned baseline model. For instance, the pruned U-Net with configuration $(1,2,3,1)$ achieves a 65\% reduction in parameters and an 18\% reduction in MACs. After finetuning it, the model gives an FAD score of 1.57 and a KL divergence of 1.678, compared with 3.95 and 2.16 for the unpruned model, respectively. Similarly, the U-Net with configuration $(1,2,1,1)$ achieves an 83\% reduction in parameters and a 39\% reduction in MACs. After finetuning, it obtains an FAD score of 1.57 and a KL divergence of 1.778, indicating that larger model compression can be achieved while maintaining generation quality comparable to, or better than, that of the unpruned baseline. The unpruned model achieves an RTF of 2.14 and requires 8.8 GB of checkpoint storage. In contrast, the pruned AudioLDM-M-Full models with U-Net configurations of (1,2,1,1) and (1,2,3,1) achieve RTFs of 1.86 and 1.88 while reducing checkpoint storage requirements to 4.4 GB and 3.2 GB, respectively.

\begin{figure}[t]
    \centering
    \includegraphics[scale=0.235]{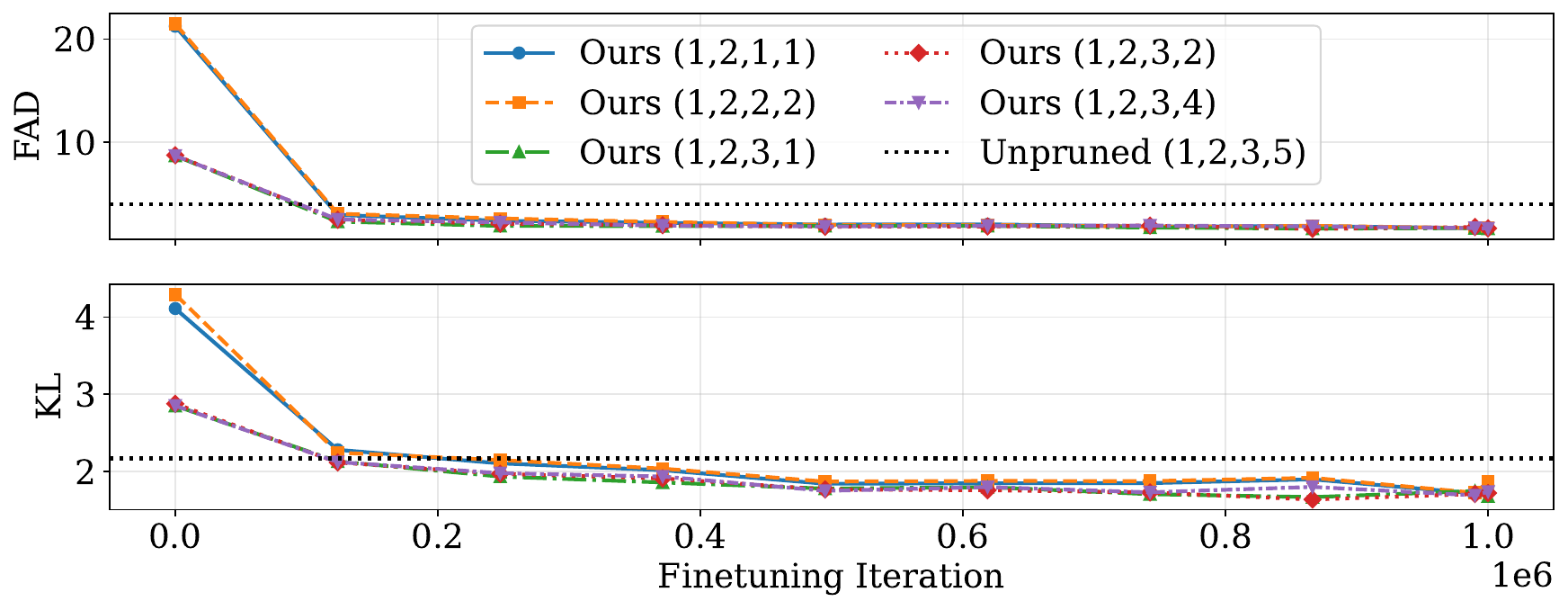}
        \vspace{-0.35cm}
    \caption{FAD and KL performance during finetuning for different $(b_1,b_2,b_3,b_4)$ configurations. }
    \label{fig: finetuning performance}
\end{figure}

Fig. \ref{fig: finetuning performance} shows performance during finetuning of the pruned models. We find that the pruned network recovers the unpruned baseline performance within 0.2M steps during finetuning. 

\begin{figure}[ht]
    \centering
    \includegraphics[scale=0.233]{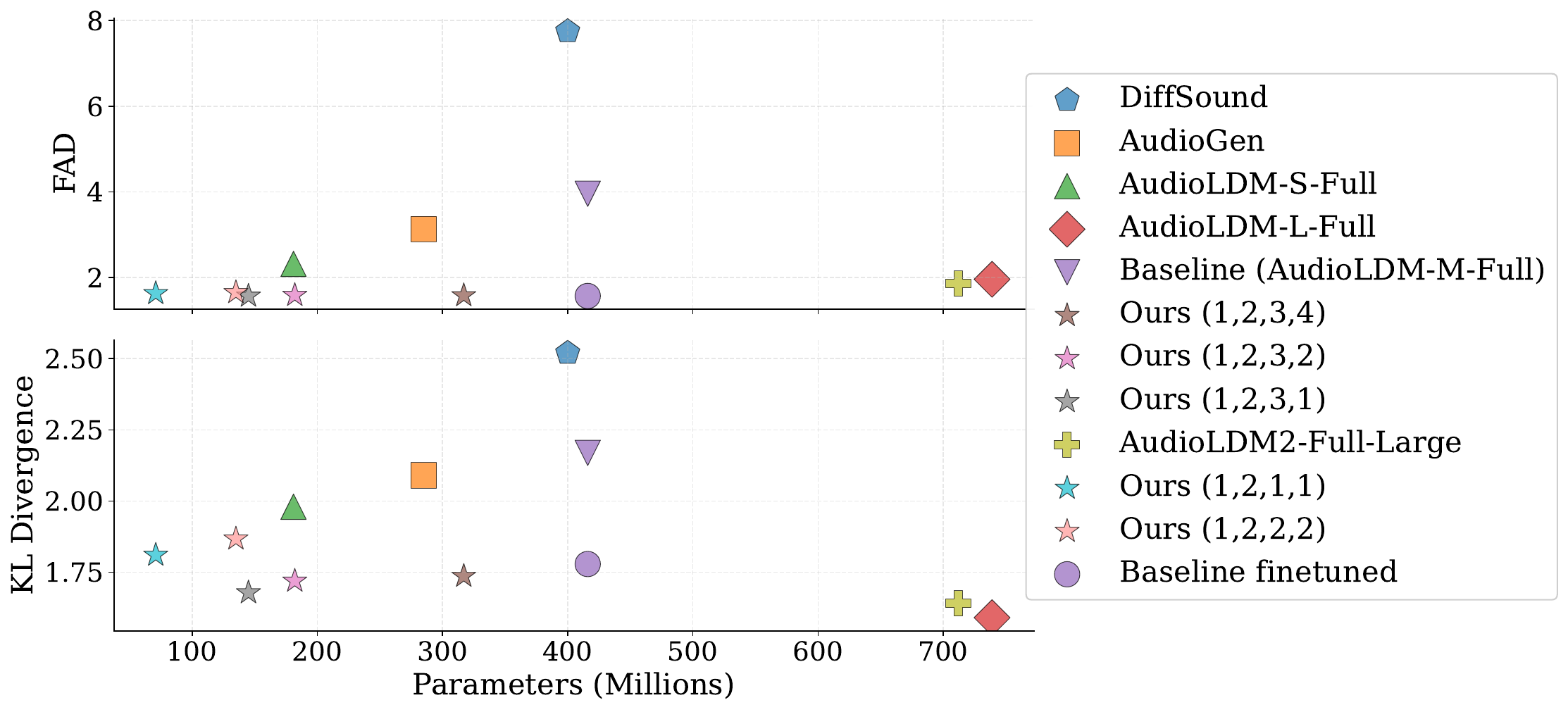}
       \vspace{-0.72cm}
    \caption{Performance comparison of our efficient pruned models with existing models. Ours ($b_1,b_2,b_3,b_4)$ denotes  channel scaling parameters across four blocks of U-Net.}
    \label{fig: scatter plot}
\end{figure}

\textbf{Comparison with other methods:} Fig. \ref{fig: scatter plot} compares performance and number of parameters across existing text-to-audio generative models.  Our proposed models achieve a good balance between the performance and the number of parameters, consistently obtaining some of the lowest KL divergence scores while using significantly fewer parameters than most of the existing models. In particular, the smaller variant, U-Net with $(1,2,1,1)$ configuration, achieves the best KL divergence values, demonstrating that high-quality audio generation can be achieved without large model sizes. The larger variant, U-Net with ($1,2,3,1$) configuration, further improve FAD, reaching performance that is competitive with or better than much larger models such as AudioLDM2-Full-Large, AudioLDM-L-Full. Although finetuning improves the performance of the unpruned baseline model, it still requires a higher parameter count and more compute in finetuning than the pruned model, even though both achieve comparable performance.

\section{Semantic quality analysis}

We utilize event-level ground-truth annotations from AudioCaps to analyze the semantic quality of audio generated by the unpruned baseline model, as well as by the pruned models with and without finetuning. For this analysis, we select the best-performing efficient AudioLDM-M-Full model with U-Net configuration $(1,2,3,1)$.

To measure semantic quality, we use a pre-trained audio neural network, PANNs \cite{kong2020panns}, to predict sound events from the generated audio. A ground-truth label is considered successfully captured if it appears among the top-10 event-level predictions produced by PANNs; otherwise, it is considered missed. We then compute the event-level capture rate (recall), as defined in Equation \ref{equ: recall}, using the ground-truth labels and the top-10 predicted sound events.


\begin{equation}
\label{equ: recall}
\mathrm{Recall}
=
\frac{\sum_{i=1}^{N} \lvert G_i \cap P_i \rvert}
     {\sum_{i=1}^{N} \lvert G_i \rvert},
\end{equation}

where $N$ is the number of audio clips,
$G_i$ denotes the set of ground-truth labels for clip $i$,
$P_i$ denotes the set of predicted labels for clip $i$,
and $\lvert\cdot\rvert$ denotes set cardinality. For the purpose of interpretable analysis, sound event labels are organized according to a hierarchical ontology, in which individual sound events are grouped into semantically related event families, as summarized in Table~\ref{tab:event_families}.


\begin{table}[t]
\centering
\caption{Event families and representative sound events.}
\label{tab:event_families}
\resizebox{\columnwidth}{!}{%
\begin{tabular}{ll}
\toprule
\textbf{Family} & \textbf{Representative events} \\
\midrule
Safety-critical & siren, alarm, gunshot, explosion \\
Mechanical & machine, engine, drill, sewing machine, industrial tools \\
Environment & rain, wind, thunder, storm, water, river, ocean, wave \\
Animals & bird, dog, cat, bark, meow, frog, pigeon, insect, rooster \\
Speech \& Human & speech, talk, voice, cry, laugh, shout, scream, cough, sneeze, breath, snore \\
Vehicles & car, bus, truck, motorcycle, train, boat, helicopter, aircraft \\
Other & labels not matching the above families \\
\bottomrule
\end{tabular}%
}
\end{table}

\begin{figure}[t]
    \centering
    \includegraphics[scale=0.253]{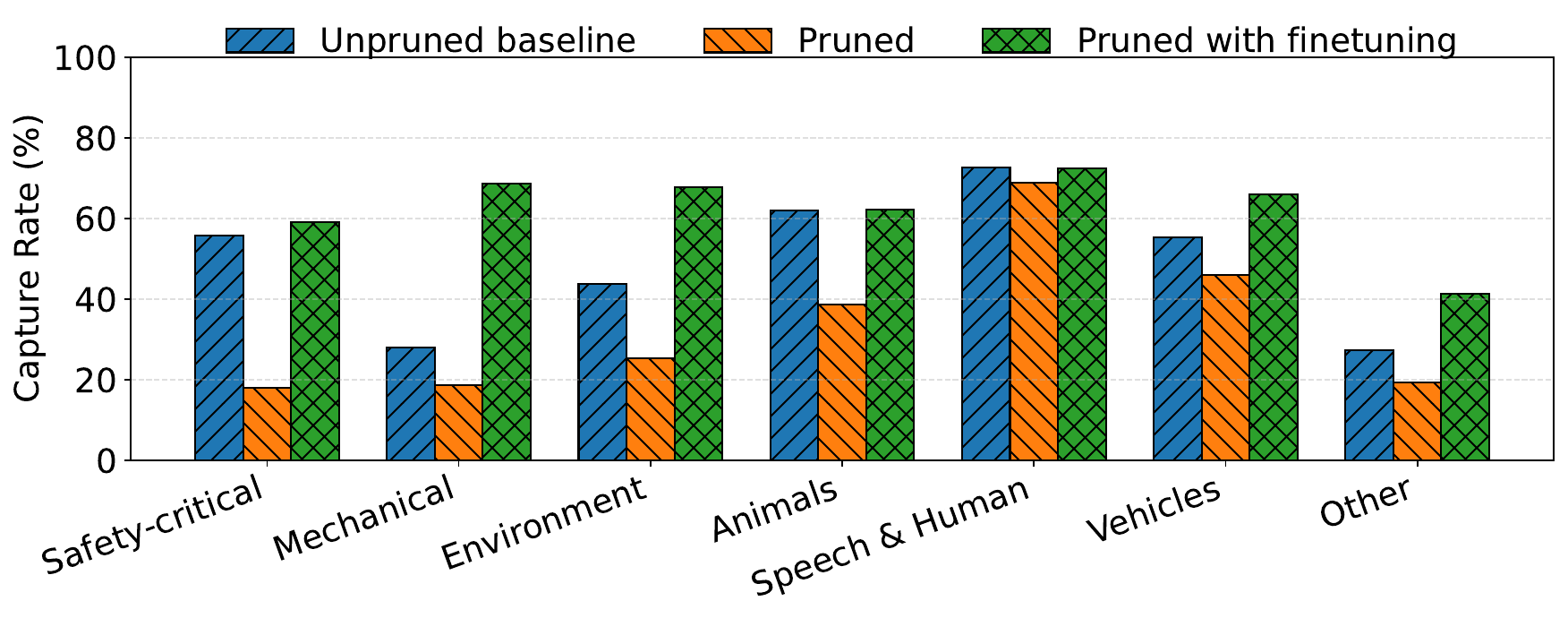}
      \vspace{-0.35cm}
    \caption{Capture rate across different sound event families  generated using unpruned, pruned and pruned with finetuning models.}
    \label{fig:barplot}
\end{figure}

\begin{figure}[ht]
    \centering
    \includegraphics[scale=0.272]{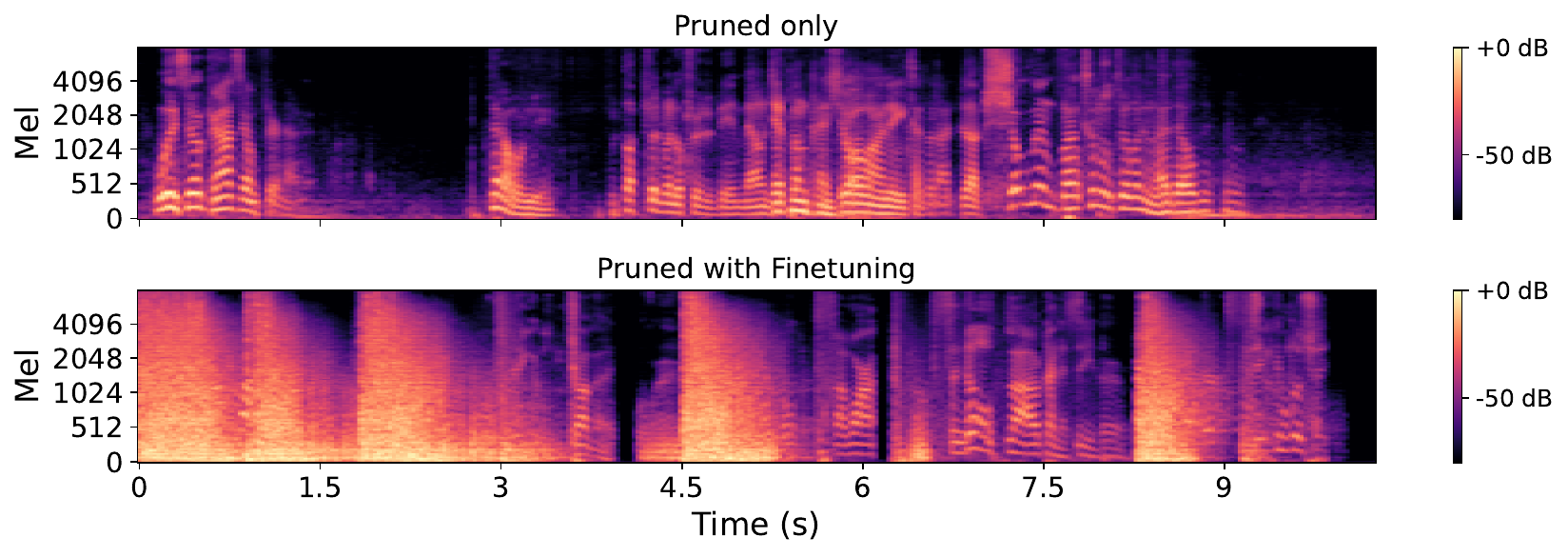}
    \vspace{-0.35cm}
    \caption{Spectrograms of audio generated using pruned models with/without finetuning given text input \enquote{Men speak with gunshots and booms}.}
    \label{fig: spectrograms examples}
\end{figure}

Fig. \ref{fig:barplot} compares capture rate among unpruned, pruned, and pruned with finetuning models. We find that model pruning leads to  degradation in capture rate across all event families. However, the extent of degradation varies by sound type. \enquote{Speech \& Human} events are the most robust to pruning, with only a small decrease in recall. 
\enquote{Vehicle} events also show moderate robustness but still experience a noticeable recall drop after pruning.
In contrast, \enquote{Animals}, \enquote{Environment}, \enquote{Mechanical}, and \enquote{Safety‑critical} events are significantly affected by pruning, with more reductions in capture rate. These events often correspond to quieter, background, or less frequent sounds, and therefore appear more sensitive to reduced model capacity. \enquote{Safety‑critical} events, despite their importance, show particularly large recall loss when pruning is applied. For instance, Fig. \ref{fig: spectrograms examples} shows spectrogram of sounds generated for an input of \enquote{Men speak with gunshots and booms} using  pruned model with and without any finetuning. Pruned model only captures the \enquote{men speaking}, on the other hand pruned model after finetuning captures all the sound events. More generated audio examples can be found here\footnote{https://arshdeep-singh-boparai.github.io/EfficientAudioLDM/}.

Finetuning the pruned model results in a similar  performance or exceeds that of the unpruned model, especially for \enquote{Vehicles}, \enquote{Environment}, \enquote{Mechanical}, and \enquote{Other} events.  Furthermore, Table \ref{tab:event_family_summary} provides number of sound events missed/recovered due to pruning and  finetuning, and sound event labels that are mostly affected due to pruning. 
Common sounds such as speech, vehicle, and bird appear frequently among those lost after pruning, suggesting that even important and frequent classes can be affected. Safety-critical sounds such as  gunshot, siren, and explosion, are being significantly impacted due to pruning. Finetuning helps to recover a large portion of the lost performance across all categories.

\begin{table}[ht]
\centering

\caption{Pruning loss and finetuning (FT) recovery by event family. Loss and recovery are measured relative to  the number of events detected by the unpruned baseline. Representative events are reported as \textit{missed/recovered} detections after pruning and FT.}
\label{tab:event_family_summary}
\normalsize
\resizebox{\columnwidth}{!}{%
\begin{tabular}{lccc p{6cm}}
\toprule
Family & Unpruned & Loss (\%) & Rec. (\%) & Representative events \\
\midrule
Safety-critical & 34 & 73.5 & 76.0 & Gunshot (8/7), Siren (7/5), Explosion (5/3) \\

Mechanical & 24 & 75.0 & 83.3 & Drill (5/4), Sewing machine (4/3), Engine (3/3) \\
Environment & 76 & 57.9 & 79.5 & Wind (6/5), Wind noise (6/6), Thunder (5/5) \\
Animals & 184 & 47.3 & 69.0 & Bird (18/13), Animal (17/15), Pets (10/9) \\
Speech \& Human & 358 & 13.1 & 74.5 & Speech (34/29), Whimper (4/1), Baby cry (4/2) \\
Vehicles & 310 & 28.4 & 75.0 & Vehicle (12/11), Car (10/9), Boat (9/9) \\
Other & 500 & 37.0 & 57.8 &  Spray (8/7), Tick-tock (8/5) \\

\bottomrule
\end{tabular}%
}
\end{table}

\section{Conclusion}



This paper applies a passive filter pruning for improving the efficiency of a diffusion-based text-to-audio generation model. By applying an $\ell_1$-norm-based pruning strategy followed by lightweight finetuning, we achieved up to 83\% parameter reduction and 39\% compute reduction while maintaining generation quality comparable to that of the original unpruned model. Furthermore, pruning was found to have a significant impact on certain sound categories, particularly safety-critical and mechanical sounds, although much of the resulting degradation could be recovered through finetuning. These findings suggest that text-to-audio diffusion models contain significant redundancy, and  pruning is an effective approach for improving model efficiency with minimal impact on generation quality. In future, we would like to apply LoRA during finetuning to further reduce the computations during finetuning, and explore one step generation methods for faster inference.

\section{Acknowledgment}
\label{sec:ack}

This work was supported by the Engineering and Physical Sciences Research Council (EPSRC)  [grant number EP/Y028805/1].  For the purpose of open access, the authors have applied a Creative Commons Attribution (CC BY) licence to any Author Accepted Manuscript version arising.


\bibliographystyle{IEEEtran}
\bibliography{refs}







\end{document}